\documentclass[aps,prb,amsmath,amssymb,superscriptaddress,floatfix,showpacs]{revtex4-2}

\def\lsim{\mathrel{\rlap{\lower4pt\hbox{\hskip1pt$\sim$}}
		\raise1pt\hbox{$<$}}}
\def\gsim{\mathrel{\rlap{\lower4pt\hbox{\hskip1pt$\sim$}}
		\raise1pt\hbox{$>$}}}

\usepackage[T3,T1]{fontenc}
\usepackage[cal=boondox,scr=boondoxo]{mathalfa}
\usepackage{subfigure}
\usepackage{mathtools}
\usepackage[mathscr]{euscript}

\usepackage{slashed}
\usepackage[hidelinks]{hyperref}
\usepackage{color}
\usepackage{colortbl}
\usepackage{amsmath, amsfonts, amssymb, amsthm}
\allowdisplaybreaks

\usepackage{bm}
\usepackage{bbm}
\usepackage{siunitx}
\usepackage{booktabs}
\makeatletter

\newcommand{\beq}{\begin{equation}}
	\newcommand{\eeq}{\end{equation}}
\newcommand{\rf}[1]{(\ref{#1})}

\def\etal{{\it et al.}}

\def\lrprt#1{\hskip-2pt\stackrel{\leftrightarrow}{\,\partial_{#1}}\hskip-2pt}

\newcommand{\half}{\tfrac12}

\newcommand{\be}{\begin{equation}}
	\newcommand{\ee}{\end{equation}}
\newcommand{\bea}{\begin{align}}
	\newcommand{\eea}{\end{align}}

\newcommand{\rhoTbIG}{7.35\ \mathrm{g\,cm}^{-3}}                 
\newcommand{\MTbIG}{947.99\ \mathrm{g\,mol}^{-1}}              
\newcommand{\nfuTbIG}{4.67\times10^{21}\ \mathrm{cm}^{-3}}      

\newcommand{\BwidthCL}{7.54\times10^{-24}\ \mathrm{GeV}}

\newcommand{\gzeroCL}{3.83\times10^{-4}\ \mathrm{GeV}^{-2}}
\newcommand{\ZetaTwoCL}{2.92\times10^{-13}\ \mathrm{GeV}}
\newcommand{\gzetaTwoCL}{1.48\times10^{7}\ \mathrm{GeV}^{-2}}

\begin{document}

\title{Torsion and Nonmetricity Constraints \\ From Neutron Spin Rotation in Polarized Matter}

\author{Sepehr Samiei}
\affiliation{
	Physics Department, Indiana University, 
	Bloomington, IN 47405, USA}

\author{Ralf Lehnert}
\affiliation{
	Physics Department, Indiana University, 
	Bloomington, IN 47405, USA}
\affiliation{
	Indiana University Center for Spacetime Symmetries, 
	Bloomington, IN 47405, USA}

\author{W.\ Michael Snow}
\affiliation{
	Physics Department, Indiana University, 
	Bloomington, IN 47405, USA}
\affiliation{
	Indiana University Center for Spacetime Symmetries, 
	Bloomington, IN 47405, USA}
\affiliation{
	Center for Exploration of Energy and Matter, 
	Indiana University,
	Bloomington, IN 47408, USA}
\affiliation{
	Quantum Science and Engineering Center, 
	Indiana University,
	Bloomington, IN 47405, USA}

\date{August 2026} 

\begin{abstract}
	Many modified-gravity frameworks employ non-Riemannian connection degrees of freedom, 
	including torsion and nonmetricity. 
	In Einstein--Cartan-type theories, 
	torsion is associated with spin density, 
	while more general metric-affine frameworks 
	can be investigated through phenomenological searches 
	for in-matter torsion and nonmetricity backgrounds. 
	This work applies an effective field theory approach 
	to characterize general couplings of electron and neutron operators 
	to spacetime torsion and nonmetricity 
	up to mass dimension six, 
	predicting novel neutron-polarization signatures 
	in the presence of an electron background. 
	Neutron spin-rotation data 
	from a polarized-electron ensemble inside a compensated ferrimagnet 
	are used to constrain for the first time 
	axially symmetric, direction-dependent in-matter torsion and nonmetricity combinations.
\end{abstract}


\maketitle

\section{Introduction}
\label{sec:intro}

Our current description of classical gravitational phenomena 
treats spacetime as a Riemannian manifold 
with geometrical features 
that are ascribed dynamical physical properties. 
In this approach, 
the gravitational field
is encoded in spacetime curvature 
and is sourced by the energy--momentum tensor. 
This physical picture 
has weathered a multitude of experimental and observational tests. 

Does spacetime curvature capture all physical phenomena 
associated with the geometry of spacetime?  
More general metric-affine structures in 4-dimensional spacetime 
may accommodate physical degrees of freedom 
not determined solely by the Riemannian curvature. 
In metric-affine gravity~\cite{bl13}, 
one extends Riemannian geometry by dropping both 
the symmetry condition 
$\Gamma^\lambda {}_{\mu\nu}-\Gamma^\lambda {}_{\nu\mu}=0$ 
on the connection coefficients 
and the metric-compatibility requirement
$D_\lambda g_{\mu\nu}=0$. 
To parametrize departures from these conditions 
one introduces two tensor fields
\begin{equation}\label{NMandTdef}
	T^\lambda{}_{\mu\nu}\equiv \Gamma^\lambda {}_{\mu\nu}-\Gamma^\lambda {}_{\nu\mu}\,,\qquad
	N_{\lambda\mu\nu}\equiv -D_\lambda g_{\mu\nu}\,,
\end{equation}
termed torsion and nonmetricity, respectively.  

Torsion arises naturally 
in gauge formulations of gravity based on local Poincar\'e symmetry, 
as emphasized in the classic work of Kibble and Sciama~\cite{Kibble1961,Sciama1964}, 
and by the need to describe spinors in curved spacetime geometries. 
The Riemann--Cartan geometry
that implements these ideas 
includes nonvanishing spacetime torsion
while maintaining zero nonmetricity $N_{\lambda\mu\nu}=0$~\cite{torsion} 
and is known as Einstein--Cartan (EC) theory~\cite{ca22}. 
More generally, 
one can treat torsion phenomenologically 
and explore the experimental limits on background torsion $T^\lambda{}_{\mu\nu}$~\cite{Overview,micro,SunSource,astro,Kaons,CL97,LHC,GravWave,GProbeB,TorsionSME,InMatter1,InMatter2,PrefFrTests}. 
Complementary theories with nonvanishing nonmetricity but $T^\lambda{}_{\mu\nu}=0$ 
have also been investigated~\cite{nm0}.
Possible couplings of ordinary matter 
to hypothetical nonmetricity fields~\cite{nm1,nm2} 
have been analyzed phenomenologically 
to constrain various $N_{\lambda\mu\nu}$ components from experimental data.

Some affine-gravity models 
only generate nonzero torsion or nonmetricity in spacetime regions 
where the relevant microscopic source density is present. 
In EC theory, 
for example, 
torsion vanishes outside regions with nonzero spin density, 
and a spin-polarized medium is therefore the appropriate source 
to search for this class of effects.
The experimental investigation of such theories 
typically requires precision spin-dependent measurements of polarized particles inside polarized matter. 
The acute technical difficulties encountered in such measurements, 
usually associated with overwhelming magnetic backgrounds 
and with the challenge of understanding dense systems of polarized particles well enough 
to isolate small deviations from strong and electroweak effects, 
have severely limited our ability to test these ideas experimentally.

Polarized slow neutron spin rotation 
can break this experimental deadlock. 
The ability of slow neutrons to penetrate macroscopic amounts of matter 
with negligible decoherence rates in certain media 
allows spin-dependent quantum amplitudes 
in a beam of polarized slow neutrons passing through matter 
to accumulate large phase shifts, 
which can be sensed with interferometric measurements, 
such as neutron polarimetry~\cite{Nico05b, Dubbers11, Pignol:2015}. 
In-matter torsion and nonmetricity fields can be sought 
if one can suppress background effects 
from the interaction of the neutron magnetic moment with magnetic fields. 

We performed the first such laboratory search several years ago~\cite{InMatter1} 
through a sensitive measurement of the neutron optical rotary power of liquid ${}^4\textrm{He}$. 
We analyzed the results with the $^{4}$He 
treated as a source of torsion and nonmetricity. 
Since the liquid ${}^4 \textrm{He}$ medium has no net spin, 
our experimental constraints on in-matter torsion and nonmetricity from this work 
were restricted to spacetime torsion and nonmetricity operators 
that yield rotation-invariant, 
or ``isotropic,'' 
terms in the nonrelativistic neutron Hamiltonian. 
In the present work, 
a polarized ferrimagnet serves as the torsion and nonmetricity source, 
selecting a preferred axis. 
The relevant effects are therefore no longer isotropic: 
they only maintain axial symmetry and can otherwise be direction dependent. 
The most physically interesting and distinct components of torsion and nonmetricity interactions, 
however, 
may be precisely these direction-dependent components of $T^\lambda{}_{\mu\nu}$ and $N_{\lambda\mu\nu}$. 
To search for these components experimentally, 
one must prepare a spin-ordered ensemble 
without creating an overwhelming background from internal magnetic fields.   

In the present work,
we report the first experimental constraints 
on axially symmetric, direction-dependent components of in-matter torsion and nonmetricity. 
We have recently solved the polarized source magnetic background issue 
by developing a ferrimagnetic compound, 
terbium iron garnet  $\textrm{Tb}_3 \textrm{Fe}_5 \textrm{O}_{12}$ (TbIG) held at its compensation temperature, 
as a polarized-electron source medium.
At its compensation temperature, 
this material possesses a net electron polarization 
with a vanishing internal magnetization. 
This source can generate or select axially symmetric torsion backgrounds in EC-type models and, 
more generally, 
provides a controlled polarized medium for phenomenological torsion and nonmetricity searches. 
As in Ref.~\cite{InMatter1}, 
polarized neutrons traversing the TbIG source 
will change their polarization state 
if torsion or nonmetricity interactions are present.  

In this paper, 
we extract limits on in-matter torsion and nonmetricity 
from our experimental bound on the neutron spin rotation angle. 
The outline of this work is as follows. 
Section~\ref{EFT} reviews our phenomenological analysis of torsion and nonmetricity. 
In Sec.~\ref{Exp}, 
we provide a brief description of our experimental setup.
Section~\ref{Res} presents and discusses our measurement results. 
In Sec.~\ref{sec:comparison_app}, 
we compare with previous work. 
Our conclusions and suggestions for future work 
are contained in Sec.~\ref{Concl}.
We adopt natural units $c=\hbar=k_B=1$ throughout unless noted otherwise. 
Our conventions for the metric signature and the Levi-Civita symbol are
$\eta^{\mu\nu}=\textrm{diag}(+,-,-,-)$
and $\epsilon^{0123}=+1$, 
respectively.

\section{Framework Basics and Phenomenology}
\label{EFT}

We adopt an effective field theory (EFT) approach to model neutrons 
moving through a medium of polarized electrons 
interacting via torsion $T^\lambda{}_{\mu\nu}$ and nonmetricity $N_{\lambda\mu\nu}$ fields. 
In Standard-Model physics on a Riemannian manifold,  
nonminimal couplings between curvature and Standard-Model fields 
are known to be contained in the EFT of gravity and matter~\cite{Drummond:1979pp,Bastianelli:2008cu}. 
It is therefore natural 
to include nonminimal $T^\lambda{}_{\mu\nu}$ and $N_{\lambda\mu\nu}$ couplings 
in our EFT approach.
Since we are interested in low-energy physics, 
we may model the neutron as a point Dirac fermion. 
Since ordinary gravitational effects are negligible, 
we may work in the flat-spacetime Minkowski limit $g^{\mu\nu}\to\eta^{\mu\nu}$. 
The free neutron and electron Lagrangians 
then take the usual form 
${\cal L}_w^0=\half\, \bar{\psi}_w \gamma^\mu i\!\lrprt\mu \psi_w - \bar{\psi}_w m_w\, \psi_w $,
where $w=e,n$ labels electron and neutron quantities,
respectively.

The construction of the EFT corrections 
${\cal L}_w^T$ and ${\cal L}_w^{N}$ to ${\cal L}_w^0$ 
is presented in Refs.~\cite{TorsionSME,nm1} 
and relies on the decomposition of both $T^\lambda{}_{\mu\nu}$ and $N_{\lambda\mu\nu}$ 
into their irreducible pieces:
\begin{align}\label{eq:Tdecomp}
	T_{\lambda\mu\nu} = 
	{}&\textstyle{\frac{1}{3}}(\eta_{\lambda\mu}T_\nu-\eta_{\lambda\nu}T_\mu)
	+\epsilon_{\mu\nu\lambda\alpha}A^\alpha
	+M_{\lambda\mu\nu}\,,\nonumber\\
	N_{\lambda\mu\nu} = {}
	{}&\tfrac{1}{18}\big[(N_1)_\mu \eta_{\nu\lambda}+(N_1)_\nu \eta_{\lambda\mu}-5(N_1)_\lambda \eta_{\mu\nu}\nonumber\\
	{}&\hspace{+11pt}-4(N_2)_\mu \eta_{\nu\lambda}-4(N_2)_\nu \eta_{\lambda\mu}+2(N_2)_\lambda \eta_{\mu\nu}\big]\nonumber\\
	{}&+S_{\lambda\mu\nu}+W_{\lambda\mu\nu}\,.
\end{align}
Explicit expressions for these irreducible components 
in terms of $T^\lambda{}_{\mu\nu}$ and $N_{\lambda\mu\nu}$ 
can be determined from Refs.~\cite{TorsionSME,nm1}, 
keeping in mind that those studies employ the opposite conventions 
for the metric and the Levi-Civita symbol.
Our notation for the mixed-symmetry piece $W_{\lambda\mu\nu}$ of $N_{\lambda\mu\nu}$ 
deviates from the conventional one 
to avoid confusion with the $M^{\lambda}{}_{\mu\nu}$ component of $T^{\lambda}{}_{\mu\nu}$. 

Up to mass dimension $d=5$, 
the torsion couplings ${\cal L}_w^{(d)T}$ 
and nonmetricity couplings ${\cal L}_w^{(d)N}$ are given by~\cite{TorsionSME,nm1}
\begin{align}\label{explL}
	{\cal L}_w^{(4)T} = 
	{}&\big[\xi_{2,w}^{(4)} T_\mu+\xi_{4,w}^{(4)} A_\mu\big]\bar{\psi}_w\gamma_5\gamma^\mu\psi_w\,,
	\nonumber\\
	{\cal L}_w^{(5)T} = 
	{}&\tfrac{i}{2}\,\xi^{(5)}_{5,w} M^{\lambda}{}_{\mu\nu}\bar{\psi}_w\sigma^{\mu\nu}\!\lrprt\lambda\psi_w
	\nonumber\\
	{}&-\tfrac{i}{2}\,\epsilon^{\kappa\lambda\mu\nu}\big[\xi_{8,w}^{(5)} T_\kappa+\xi_{9,w}^{(5)} A_\kappa\big]\bar{\psi}_w\sigma_{\mu\nu}\!\lrprt\lambda\psi_w\,,
	\nonumber\\
	{\cal L}_w^{(4)N} = 
	{}&\zeta_{2,w}^{(4)}\left(N_1\right)_\mu \bar{\psi}_w \gamma_5 \gamma^\mu \psi_w 
	+\zeta_{4,w}^{(4)}\left(N_2\right)_\mu \bar{\psi}_w \gamma_5 \gamma^\mu \psi_w\,,
	\nonumber\\
	\mathcal{L}_w^{(5)N}= 
	{}&-\tfrac{i}{4}  \zeta_{5,w}^{(5)} W_{\mu \nu}{ }^\rho \bar{\psi}_w \sigma^{\mu \nu} \!\lrprt\rho \psi_w 
	\nonumber\\
	{}&+\tfrac{i}{8}  \zeta_{6,w}^{(5)} \epsilon_{\kappa \lambda \mu \nu} W^{\kappa \lambda \rho} \bar{\psi}_w \sigma^{\mu \nu} \!\lrprt\rho \psi_w 
	\nonumber\\
	{}&-\tfrac{i}{4}  \zeta_{9,w}^{(5)} \epsilon^{\lambda \mu \nu \rho}\left(N_1\right)_\lambda \bar{\psi}_w \sigma_{\mu \nu} \!\lrprt\rho \psi_w 
	\nonumber\\
	{}&-\tfrac{i}{4}  \zeta_{10,w}^{(5)} \epsilon^{\lambda \mu \nu \rho}\left(N_2\right)_\lambda \bar{\psi}_w \sigma_{\mu \nu} \!\lrprt\rho \psi_w\,.
\end{align}
Here,
$\xi^{(d)}_{l,w}$ and $\zeta^{(d)}_{l,w}$ are real-valued couplings 
of mass dimension $d-4$
determined by the underlying theory. 
Terms that are unobservable at leading order in the present context
have been omitted~\cite{TorsionSME,nm1}.
The totally symmetric piece $S_{\lambda\mu\nu}$ of $N_{\lambda\mu\nu}$
only contributes at $d=6$ and higher:
\begin{align}\label{explL6}
	{\cal L}_w^{(6)N} \supset 
	{}&-\tfrac{1}{4} \zeta_{1,w}^{(6)} S_\lambda{ }^{\mu \nu} \bar{\psi}_w \gamma^\lambda \partial_\mu \partial_\nu \psi_w+\text { h.c. }
	\nonumber\\
	{}&-\tfrac{1}{4} \zeta_{2,w}^{(6)} S_\lambda{ }^{\mu \nu} \bar{\psi}_w \gamma_5 \gamma^\lambda \partial_\mu \partial_\nu \psi_w+\text { h.c. }
\end{align}
The analyses in this work 
are based on both Eqs.~\eqref{explL} and~\eqref{explL6}.
We will make no assumptions 
regarding the dynamics of the torsion and nonmetricity fields.

The Lagrangian contributions~(\ref{explL}) and~(\ref{explL6}) 
can now be used to extract the key torsion and nonmetricity signatures 
for our experiment. 
The polarized electrons in the TbIG 
constitute the spin-polarized source medium 
and define the preferred axis for possible in-matter torsion backgrounds in EC-type models, 
and more generally for phenomenological torsion and nonmetricity fields 
seen by the polarized neutron beam 
that moves through the sample.
Although the electron couplings $\xi^{(d)}_{l,e}, \zeta^{(d)}_{l,e}$ 
and the torsion and nonmetricity equations of motion are unknown, 
we can constrain torsion and nonmetricity 
through their contribution to the neutron optical potential of matter. 
We average both $T^\lambda{}_{\mu\nu}(\bm r) \to \langle T^\lambda{}_{\mu\nu}\rangle$ 
and $N_{\lambda\mu\nu}(\bm r) \to \langle N_{\lambda\mu\nu}\rangle$ 
over the sample and assume 
that the primary physical effects on neutron propagation 
are governed by these two spacetime-independent averages. 
This assumption is supported 
by general results in nonrelativistic scattering theory, 
which relate the neutron forward scattering amplitude 
to the neutron optical potential of the matter.

The structures of $\langle T^\lambda{}_{\mu\nu}\rangle$ 
and $\langle N_{\lambda\mu\nu}\rangle$ 
relevant for our analysis 
depend on the experiment's geometry and setup. 
The TbIG in the measurement was prepared 
with a net electron polarization 
along the axis of the disk-shaped sample. 
This implies that $\langle T^\lambda{}_{\mu\nu}\rangle \to \langle T_\parallel^\lambda{}_{\mu\nu}\rangle$ 
and $\langle N^\lambda{}_{\mu\nu}\rangle \to \langle N_\parallel^\lambda{}_{\mu\nu}\rangle$ 
exhibit rotational invariance about the electron-polarization axis,
where the $\parallel$ subscript denotes this axial symmetry. 
To characterize the axially symmetric pieces  
$\langle T_\parallel^\lambda{}_{\mu\nu}\rangle$ 
and $\langle N_\parallel^\lambda{}_{\mu\nu}\rangle$,
we choose Cartesian coordinates $j=1,2,3$
with the 3-axis aligned with the net electron polarization. 
We find that $\langle T_\parallel^\lambda{}_{\mu\nu}\rangle$ 
contains the eight independent components
$A_0$,	
$A_3$, 
$T_0$, 
$T_3$, 
$M^{0}{}_{30}$,
$M^{3}{}_{30}$,
$\epsilon^{3kl}M^{0}{}_{kl}$, and
$\epsilon^{3kl}M^{3}{}_{kl}$. 
Likewise,
$\langle N_\parallel^\lambda{}_{\mu\nu}\rangle$ 
consists of the twelve independent components 
$(N_1)_0$,	
$(N_1)_3$, 
$(N_2)_0$, 
$(N_2)_3$, 
$S_{000}$,
$S_{300}$,
$S_{330}$,
$S_{333}$,
$W_{300}$,
$W_{330}$,
$\epsilon^{3kl}W_{kl0}$, and
$\epsilon^{3kl}W_{kl3}$. 
In these expressions, 
$\epsilon^{jkl}=\epsilon^{0jkl}$,
Latin indices $j,k,l=1,2,3$ are purely spatial,
and repeated indices are summed over.
To arrive at this counting,
various non-axiality, 
tracelessness, 
and cyclicity properties 
of the individual irreducible pieces 
have been used~\cite{TorsionSME,nm1}.
Appendix~\ref{axsymback} gives explicit expressions 
for $\langle T_\parallel^\lambda{}_{\mu\nu}\rangle$ 
and $\langle N_\parallel^\lambda{}_{\mu\nu}\rangle$ 
in terms of these 20 independent components.

A slow neutron moving in this medium 
that couples to torsion and nonmetricity through Eqs.~(\ref{explL}) and~(\ref{explL6}) 
will sense these $\langle T_\parallel^\lambda{}_{\mu\nu}\rangle$ 
and $\langle N_\parallel^\lambda{}_{\mu\nu}\rangle$ backgrounds. 
The leading effects on the neutron spin, 
obtained from a Foldy--Wouthuysen reduction~\cite{nonrel_limit} of the neutron dynamics~(\ref{explL}) and~(\ref{explL6}), 
can be expressed as Hamiltonian corrections
\begin{equation}
	\delta H_{\rm spin}
	=\delta H_{\rm spin}^{(T)}+\delta H_{\rm spin}^{(N)}
	=\delta\bm\xi\!\cdot\!\bm\sigma
	+\delta\bm\zeta\!\cdot\!\bm\sigma\,,
	\label{Hcorrection}
\end{equation}
where $\bm{\sigma}$ are the usual three Pauli matrices, 
and where we have discarded terms proportional to the identity matrix, 
as they leave neutron-spin motion unaffected.
The background torsion and nonmetricity contributions are contained in $\delta\bm\xi$ and $\delta\bm\zeta$;
these coefficient vectors can be constructed from the 20 independent components 
of $\langle T_\parallel^\lambda{}_{\mu\nu}\rangle$ 
and $\langle N_\parallel^\lambda{}_{\mu\nu}\rangle$,
the neutron momentum $\bm p$, 
and the neutron mass $m\equiv m_n$.
The nonrelativistic expansions
\begin{align}\label{eq:FW_expansions}
	\delta\bm\xi
	& =\delta\bm\xi_0+\beta\,\delta\bm\xi_1
	+\beta^2\delta\bm\xi_2+\mathcal O(\beta^3)\,,
	\nonumber\\
	\delta\bm\zeta
	& =\delta\bm\zeta_0+\beta\,\delta\bm\zeta_1
	+\beta^2\delta\bm\zeta_2+\beta^3\delta\bm\zeta_3
	+\mathcal O(\beta^4)\,,
\end{align}
where $\beta=|\bm p|/m$, 
are explicitly given 
in terms of torsion and nonmetricity in Appendix~\ref{axsymback}.

The spin Hamiltonian~(\ref{Hcorrection}) 
contains all 20 independent axial-symmetry torsion and nonmetricity parameters. 
However, 
the details of the experimental setup 
may only provide access to a subset of these.
The present experiment,
for example,
involves a neutron beam 
aligned with the axial 3-direction,
so that $\bm{\hat{p}}\to \bm{\hat{z}}$ and thus $\bm{p_\varphi}\to \bm{0}$ 
in Eqs.~(\ref{xi_Def}) and (\ref{zeta_Def}). 
This eliminates $M^0{}_{30}$, $M^3{}_{30}$, $W_{300}$, and $W_{330}$ 
from the spin Hamiltonian.
The coefficient vectors $\delta\bm{\xi}$ and $\delta\bm{\zeta}$ 
only retain 3-components
and both $\delta \bm{\xi_2},\,\delta \bm{\xi_3}\to \bm{0}$ in this limit;
by contrast, 
$\bm{\delta\zeta_3}$ need not vanish:
\begin{alignat}{2}\label{eq:axial_energy_coefficients}
	\delta\bm{\xi}
	&=(\delta\xi)_3\bm{\hat{z}}
	&&=\big[(\delta\xi_0)_3+(\delta\xi_1)_3\beta\big]\bm{\hat{z}}\,,
	\nonumber\\
	\delta\bm{\zeta}
	&=(\delta\zeta)_3\bm{\hat{z}}
	&&=\big[(\delta\zeta_0)_3+(\delta\zeta_1)_3\beta+(\delta\zeta_2)_3\beta^2+(\delta\zeta_3)_3\beta^3\big]\bm{\hat{z}}\,.
\end{alignat}
Spin evolution of a transversely polarized neutron inside our ferrimagnet
via $U(t)=\exp(-i\hspace{1pt}\delta H_{\rm spin}\hspace{1pt}t )$ 
then yields
\begin{equation}\label{eq:torsion_axial_one_orientation}
	\phi_+=-2\big[(\delta\xi)_3+(\delta\zeta)_3\big]\frac{L}{v}
\end{equation}
for the rotation angle $\phi_+$ of the neutron's spin 
in the plane orthogonal to its momentum. 
Here, 
$L$ is the length of the neutron's path inside the ferrimagnet 
and $v$ its speed.

For practical reasons, 
our experimental observable is constructed 
as an asymmetry between $\phi_+$ and $\phi_-$ 
per unit neutron-travel length inside the ferrimagnet.
Here, 
$\phi_-$ is the neutron's spin rotation angle
measured via the same procedure  
but with the ferrimagnet rotated $180^\circ$ 
about a vertical axis,
i.e.,
with the net ferrimagnetic electron spin density reversed
$\bm S_e\to-\bm S_e$. 
In effect,
our measured observable is thus the neutron rotary power
\begin{align}\label{eq:half_difference_rotation}
	\frac{d}{dL}\phi
	&\equiv\frac{d}{dL}\frac{1}{2}(\phi_+-\phi_-)
	\nonumber\\
	&=-\frac{2}{v}\,\Big[(\delta\xi_0)_3+(\delta\zeta_0)_3+(\delta\zeta_2)_3\beta^2\Big]\,.	
\end{align}
To arrive at this expression,
we used the fact 
that terms with even powers of $\beta$ 
change sign under the ferrimagnet's rotation
while odd powers remain unaffected.
This may,
for example, 
be established 
from the rotation properties 
of the 4-vectors $\tau^\mu\to\tau^\mu$ and $a^\mu\to -a^\mu$ in Eqs.~(\ref{isoT}) and~(\ref{isoN}),	
which provide the basis for the expansion of  
$\langle T_\parallel^\lambda{}_{\mu\nu}\rangle$ 
and 
$\langle N_\parallel^\lambda{}_{\mu\nu}\rangle$. 
Thus,
terms in $\phi$ with odd powers of $\beta$ cancel.
The choice of this observable therefore 
entails a further reduction of the number 
of accessible torsion and nonmetricity parameters:
only components with an odd number of the spatial index $j=3$ contribute.

\section{Experimental Setup and Results}
\label{Exp}

The results used below 
for the anisotropic torsion and nonmetricity constraints 
were collected in a series of measurements 
at two neutron user facilities at Oak Ridge National Laboratory: 
the Spallation Neutron Source and the High Flux Isotope Reactor. 
The experiment consists of a measurement 
of the spin rotation angle per unit length of transversely polarized neutrons 
passing through the dense ensemble of polarized electrons in the TbIG ferrimagnet. 
A ferrimagnet~\cite{ferri1,ferri2} possesses two magnetic sublattices (associated here with iron and terbium ions) whose magnetic moments anti-align. Ferrimagnets often possess a compensation temperature $T_{c}$ where the magnetic moments of these two anti-aligned sublattices cancel, producing zero internal magnetization. The electron polarization, however, does not cancel: since the magnetic moments of iron and terbium ions possess different relative contributions from the orbital and spin magnetic moments, a net electron spin density is present inside the medium whose magnitude can be determined from the known magnetic properties of the ions.
If one can maintain the material at $T_{c}$, 
which can be determined experimentally by external magnetometry, 
systematic errors 
from the coupling of internal magnetic fields to the neutron magnetic moment 
are suppressed.

We performed an extensive series of measurements 
using SQUID magnetometry, 
x-ray diffraction, 
neutron small-angle scattering, 
neutron spin-echo spectroscopy, 
and polarized neutron imaging 
to confirm our understanding of the magnetic dynamics of this ferrimagnetic material 
and determine $T_{c}$~\cite{hughes_polarized_2025}. 
We have recently measured a TbIG spin polarization of
$0.55(5)\,\mu_B$ per formula unit, 
where one formula unit denotes $\mathrm{Tb}_3\mathrm{Fe}_5\mathrm{O}_{12}$.
This measured polarization agrees with the value 
inferred from a mean-field model 
applied to TbIG~\cite{BaxterPrivate2026}. 
Using $\rho_{\rm TbIG}=\rhoTbIG$ and $M_{\rm TbIG}=\MTbIG$, 
the formula-unit density is $n_{\rm f.u.}=\rho_{\rm TbIG}N_A/M_{\rm TbIG}=\nfuTbIG$, 
so this value gives 
\begin{equation}\label{eq:ns_clean}
	n_s=(0.55\pm0.05)n_{\rm f.u.}=(2.57\pm0.23)\times10^{21}\ \mathrm{cm}^{-3}	
\end{equation}
for the  effective polarized-electron density.

Since the electron spin orientation is rigidly fixed inside the material, 
the measurement is conducted 
by mechanically rotating the sample 
by $180^\circ$ about a vertical axis 
to reverse the direction of the electron spin 
relative to the neutron momentum. 
Our measurement of the neutron spin rotary power in terbium iron garnet held at $T_{c}$, 
denoted $d\bar{\phi}_{F5}/dz$ in Ref.~\cite{Mulkey2026}, 
is 
\begin{equation}\label{exp_rot_power}
	\left.\frac{d\phi}{dL}\right|_{\rm exp}\!\!
	=[0.41\pm6.30\textrm{ (stat.)}\pm4.4\textrm{ (sys.)}]\times10^{-3}\frac{\textrm{rad}}{\textrm{m}}\,, 
\end{equation}
consistent with zero. 
The target thickness was $L_{2024}=1.00\pm0.02$ cm. 
The CG-1D neutron beamline 
has an average neutron wavelength of $\lambda_n=2.6$~\AA~\cite{iverson_flux_2024},
or
\begin{equation}\label{eq:kinematic_factors_clean}
	\beta=5.08\times10^{-6}\,,
\end{equation}
which corresponds to an average nonrelativistic speed 
of about $v=1520\,\textrm{m}\,\textrm{s}^{-1}$.

\section{Anisotropic Torsion and Nonmetricity Bounds}
\label{Res}

Combining the theoretical result~(\ref{eq:half_difference_rotation}) 
with the corresponding measurement~(\ref{exp_rot_power}) 
and the neutron speed~(\ref{eq:kinematic_factors_clean})
places experimental constraints 
on a combination of in-matter torsion and nonmetricity components. 
Even though both torsion and nonmetricity may in general be present simultaneously,
we proceed below by considering scenarios 
in which either torsion or nonmetricity is present. 
Barring fortuitous cancellations, 
this approach has the benefit 
of providing individual sensitivity levels 
for each of these two distinct metric-affine structures 
beyond Riemannian geometry. 

For the 95\% C.L.\ limits quoted below, we treat the statistical and systematic
uncertainties in Eq.~\eqref{exp_rot_power} as independent 
and combine them in quadrature,
\[
\sigma_{\rm tot}=\sqrt{(6.30)^2+(4.4)^2}\times10^{-3}
=7.68\times10^{-3}\ \mathrm{rad/m}.
\]
The corresponding two-sided, null-centered 95\% C.L.\ sensitivity is
\(1.96\,\sigma_{\rm tot}=1.51\times10^{-2}\ \mathrm{rad/m}\),
which gives the spin-energy sensitivity \(B_{95}=\BwidthCL\). 
The measured central value is only
\(0.053\,\sigma_{\rm tot}\) from zero.

We begin by disregarding the nonmetricity terms
$(\delta\zeta)_3=(\delta\zeta_0)_3+(\delta\zeta_2)_3\beta^2$
and focus solely on torsion 
\begin{align}\label{eq:torsion_parameters}
	(\delta\xi)_3 =(\delta\xi_0)_3	
\end{align}
with
\begin{align}\label{eq:torsion_esd_result}
	(\delta\xi_0)_3 =
	& -m\xi_5^{(5)}\epsilon^{3kl}M^{0}{}_{kl}
	\nonumber\\
	& -(2m\xi_9^{(5)}-\xi_4^{(4)})A_3
	-(2m\xi_8^{(5)}-\xi_2^{(4)})T_3\,.
\end{align}
Our analysis then implies the bound
\begin{equation}\label{eq:torsion_bound}
	\big|
	(\delta\xi_0)_3
	\big|
	\le \BwidthCL
\end{equation}
on torsion components 
generated by the net electron polarization
inside the ferrimagnet.
With the statistical and systematic uncertainties combined in quadrature as described above,
this bound holds at $95\%$ C.L.
Normalization with respect to the electron spin density~(\ref{eq:ns_clean}),
\begin{equation}\label{eq:torsion_source_normalized_bound}
	\big|
	(\delta\xi_0)_3
	\big|\,n_s^{-1}
	\le \gzeroCL\,,
\end{equation}
provides a measure quantifying this sensitivity 
relative to the strength of the putative torsion source. 

Next, 
we ignore torsion $(\delta\xi)_3=(\delta\xi_0)_3$ 
and only consider nonmetricity
\begin{equation}\label{eq:nm_parameters}
	(\delta\zeta)_3 =(\delta\zeta_0)_3+(\delta\zeta_2)_3\beta^2
\end{equation}	
with 
\begin{align}
	(\delta\zeta_0)_3
	&=(\zeta_2^{(4)}-m\zeta_9^{(5)})(N_1)_3
	+(\zeta_4^{(4)}-m\zeta_{10}^{(5)})(N_2)_3
	\nonumber\\
	&\quad
	+\tfrac{1}{2}\zeta_2^{(6)}m^2 S_{300}
	+\tfrac{1}{2}\zeta_5^{(5)}m\epsilon^{3jk}W_{jk0}\,,
	\label{eq:zeta0_esd_def}\\
	(\delta\zeta_2)_3
	&=\tfrac{1}{2}\zeta_2^{(6)}m^2(3S_{300}+S_{333})\,.
	\label{eq:zeta2_esd_def}
\end{align}
For this case,
we obtain analogously
\begin{equation}
	\big|
	(\delta\zeta)_3
	\big|
	\le \BwidthCL
	\label{eq:nonmetricity_bound}
\end{equation}
at $95\%$ C.L.\ on nonmetricity components 
produced by the net electron spin density
inside the ferrimagnet.
Because $(\delta\zeta_0)_3$ and $(\delta\zeta_2)_3$ enter the same observable,
the individual sensitivities quoted below are single-coefficient limits, 
obtained by setting the other surviving coefficient to zero. 
Due to the $\beta^2$ suppression of $(\delta\zeta_2)_3$,
the individual sensitivities to $(\delta\zeta_0)_3$ and $(\delta\zeta_2)_3$ differ:
\begin{align}
	\left|(\delta\zeta_0)_3\right|
	&\le \BwidthCL\,,
	\label{eq:nonmet_energy_zeta0}\\
	\left|(\delta\zeta_2)_3\right|
	&\le\ZetaTwoCL\,.
	\label{eq:nonmet_energy_zeta2}
\end{align}
Normalizing relative to $n_s$,
\begin{align}
	\left|(\delta\zeta)_3\right|\,n_s^{-1}
	&\le\gzeroCL\,,
	\label{eq:nonmet_norm_g}\\
	\left|(\delta\zeta_0)_3\right|\,n_s^{-1}
	&\le\gzeroCL\,,
	\label{eq:nonmet_norm_g0}\\
	\left|(\delta\zeta_2)_3\right|\,n_s^{-1}
	&\le\gzetaTwoCL\,,
	\label{eq:nonmet_norm_g2}
\end{align}
again yields the corresponding 
sensitivities relative to the source of nonmetricity.

\section{Comparison with Related Work}
\label{sec:comparison_app}

The asymmetry (\ref{eq:half_difference_rotation}) formed 
using the ferrimagnet rotation 
is only sensitive to anisotropic physics
and by design suppresses isotropic effects. 
As a result, 
the present measurements 
constrain exclusively anisotropic torsion and nonmetricity terms
generated by the net electron polarization 
inside the source.
Our previous rotary-power experiment~\cite{InMatter1},
on the other hand,
involved an unpolarized liquid-$^4$He source 
and its analysis proceeded without constructing such a rotational asymmetry. 
These previous investigations were therefore sensitive 
solely to fully isotropic torsion and nonmetricity components 
and resulted in constraints on 
$|(2m\xi^{(5)}_9-\xi^{(4)}_4)A^0+(2m\xi^{(5)}_8-\xi^{(4)}_2)T^0|$
at the $\mathcal O(10^{-22})\,\mathrm{GeV}$ level~\cite{InMatter1}
and on $|2(\zeta^{(4)}_2-m\zeta^{(5)}_9)(N_1)_0+2(\zeta^{(4)}_4-m\zeta^{(5)}_{10})(N_2)_0+m^2\zeta^{(6)}_2 S_{000}|$ 
also at the $\mathcal O(10^{-22})\,\mathrm{GeV}$ level~\cite{nm2}. 
It is apparent 
that the present work fully complements these past results.
We also note 
that our work provides constraints different from those 
derived by reinterpreting bounds on CPT and Lorentz violation 
within the Standard-Model Extension~\cite{TorsionSME,nm1,sme,DataTables}, 
which were based on spatially separated sources and probes.

Next, 
we compare our measurement 
to the specific case of minimal EC theory,
which is recovered in the limit $\xi^{(4)}_4\to3/4$ 
with all other $\xi$ and $\zeta$ couplings vanishing.
Only the axial piece of torsion 
couples to spin-$\frac{1}{2}$ fermions in this theory
and obeys a purely algebraic equation of motion:
\be\label{eq:EC_torsion}
A^{\mu}=4\pi G\,J_5^\mu \,,
\ee
where $G$ is the gravitational coupling 
and $J^\mu_5$ the chiral current of a fermionic source.
To compare this theory to our measurement
we introduce a dimensionless factor $\chi$ into Eq.~(\ref{eq:EC_torsion})
by replacing $G\to \chi G$.
Multiplicative deviations from EC theory
are then parametrized by deviations of $\chi$ from unity.

We continue by considering the 3-component 
of this $\chi$-modified EC Eq.~(\ref{eq:EC_torsion}):
\be\label{eq:EC_torsion_chi}
\chi=\frac{A^{3}}{4\pi G\,J_5^3}\,.
\ee
We may estimate $J^3_5\simeq n_s$
using our measured spin density~(\ref{eq:ns_clean}),
and in the EC limit,
the bound~(\ref{eq:torsion_esd_result})
reduces to a numerical limit on $A_3$.
Using $\xi^{(4)}_4=3/4$ gives
$|A_3|\le (4/3)\BwidthCL$, and therefore
\be\label{eq:EC_torsion_chi_limit}
|\chi|\lsim 6.0\times10^{33}.
\ee
The uncertainty in $n_s$ changes this estimate by approximately $9\%$
and does not affect the conclusion.

Reference~\cite{InMatter2} considered the corresponding minimal
EC source--probe configuration, in which the axial current
of spin-polarized matter sources torsion and produces a spin-dependent
effect on a neutron propagating through the medium.  The present	measurement experimentally realizes this general configuration using a
compensated ferrimagnetic source and constrains the overall EC scaling
factor $\chi$ introduced above.

\section{Conclusions and Outlook}
\label{Concl}

This work presents anisotropic in-matter torsion and nonmetricity bounds 
from neutron spin rotation in a compensated ferrimagnet. 
The essential experimental operation 
is reversal of the ferrimagnetic electron spin density. 
The measured rotary power must therefore be compared 
with the electron-spin-dependent Hamiltonian component 
defined in Eq.~\eqref{eq:half_difference_rotation}. 
In the present axial geometry, 
this projection retains the torsion coefficient $(\delta\xi_0)_3$ 
and the nonmetricity combination $(\delta\zeta_0)_3+\beta^2(\delta\zeta_2)_3$. 
Terms even under electron-spin reversal 
appear in the Hamiltonian for one source orientation 
but cancel in the measured difference.

The calculations presented here 
can be applied to other experimental configurations. 
We are analyzing data from a measurement 
in which the electron spin is transverse to both 
the neutron momentum and neutron spin; 
this geometry can access different torsion and nonmetricity coefficients. 
We are also analyzing polarized x-ray data 
designed to measure the internal electron polarization density directly. 
Measurements as a function of neutron momentum 
would provide additional kinematic separation of the surviving operators. 
Further improvements could follow 
from better target quality, 
temperature uniformity and control, 
and more sensitive magnetometry 
to establish $T_c$ with higher precision. 
A dedicated measurement at an intense slow-neutron beamline 
could then improve the sensitivity 
by two to three orders of magnitude.

A broader theory connecting the polarized-electron density, 
other Standard-Model fermion bilinears in the ferrimagnet, 
and the induced in-matter torsion or nonmetricity tensors 
is a separate source-model problem. 
The present analysis instead treats the averaged in-matter tensors phenomenologically, 
transforms them under the experimentally realized electron-spin reversal, 
and retains only the Hamiltonian terms 
that can be compared directly with the measured rotary power.

\section{Acknowledgments}
\label{sec:ack}

We thank Yuri Bonder for comments 
and David Baxter for providing information prior to publication 
of the electron spin density 
measured using polarized x-ray scattering.

S.\ Samiei, R.\ Lehnert, and W.M.\ Snow acknowledge support 
from US National Science Foundation grant PHY-2209481 
and from the Indiana University Center for Spacetime Symmetries.

This research used data from the High Flux Isotope Reactor, 
a DOE Office of Science User Facility 
operated by the Oak Ridge National Laboratory. 
The beam time was allocated to MARS on proposal numbers IPTS-30635 and IPTS-32059.

We acknowledge the Indiana University Gateway Center in Mexico City 
and the Instituto de Ciencias Nucleares, 
located on the main campus of the Universidad Nacional Autónoma de México (UNAM), 
for support of the Tests Of Relativity with Spin InteractiONs (TORSION) workshop 
held in Mexico City on May 18--20, 2026 
and organized jointly with IUCSS and UNAM, 
where some of the scientific implications of this work were discussed.

\appendix

\section{\boldmath Axially Symmetric $T^{\kappa}{}_{\lambda\mu}$ and $N_{\kappa\lambda\mu}$}
\label{axsymback}

We use one fixed Cartesian coordinate system 
for both source orientations. 
In the positive orientation, 
let $a^\mu=(0,0,0,1)$ be a purely spacelike unit vector, 
chosen to point along the net ferrimagnetic electron spin density, 
and let $\tau^\mu=(1,0,0,0)$ be a purely timelike unit vector. 
The vector $a^\mu$ specifies only the axial direction 
and is dimensionless; 
the physical electron spin density has magnitude $n_s$ and reverses as $a^\mu\to-a^\mu$.
In the axial-symmetry limit,
we then have for the irreducible pieces of $\langle T_\parallel^\lambda{}_{\mu\nu}\rangle$ 
\begin{align}\label{isoT}
	T_\mu =
	{}& T_0\tau_\mu + T_3 a_\mu \,,
	\nonumber\\
	A_\mu =
	{}& A_0\tau_\mu + A_3 a_\mu \,,
	\nonumber\\
	M_{\lambda\mu\nu} =
	{}&-\tfrac{1}{2}M^{0}{}_{30}\left(\eta_{\lambda[\mu } a_{\nu]} 
	+3 \tau _{\lambda } 
	a_{[\mu}\tau_{\nu]}\right)
	\nonumber\\
	{}&-\tfrac{1}{2}M^{3}{}_{30}\left(\eta_{\lambda[\mu } \tau_{\nu]} 
	+3 a _{\lambda } 
	a_{[\mu}\tau_{\nu]}\right)
	\nonumber\\
	{}&-\tfrac{1}{4}\epsilon^{3kl}M^{0}{}_{kl}
	\left(2 \tau _{\lambda } \epsilon _{\mu \nu \alpha \beta }-\tau _{[\mu } \epsilon _{\nu] \lambda \alpha \beta }\right)\tau ^{\alpha } a^{\beta }
	\nonumber\\
	{}&-\tfrac{1}{4} \epsilon^{3kl}M^{3}{}_{kl} \left(2 a_{\lambda } \epsilon _{\mu \nu \alpha \beta }-a_{[\mu } \epsilon _{\nu]
		\lambda \alpha \beta }\right)\tau ^{\alpha } a^{\beta }\,,
\end{align}
where a square bracket denotes antisymmetrization 
in the enclosed indices,
e.g., 
$A_{[\mu}B_{\nu]}\coloneqq A_{\mu}B_{\nu}-A_{\nu}B_{\mu}$.
Likewise, the axially symmetric $\langle N_\parallel^\lambda{}_{\mu\nu}\rangle$ 
can be expressed in terms of
\begin{align}\label{isoN}
	(N_1)_\mu =
	{}& (N_1)_0\tau_\mu + (N_1)_3 a_\mu \,,
	\nonumber\\
	(N_2)_\mu =
	{}& (N_2)_0\tau_\mu + (N_2)_3 a_\mu \,,
	\nonumber\\
	S_{\lambda\mu\nu} =
	{}& \tfrac{1}{2}S_{000}\left(5 \tau_{\lambda } \tau_{\mu } \tau_{\nu }- \left\{ \tau a a \right\} _{\lambda\mu\nu} -\left\{\tau\eta\right\}_{\lambda\mu\nu}\right) 
	\nonumber\\
	{}& +\tfrac{1}{2}S_{300}\left(3 a_{\lambda } a_{\mu } a_{\nu }- 3\left\{ \tau \tau a \right\} _{\lambda\mu\nu} +\left\{a\eta\right\}_{\lambda\mu\nu}\right) 
	\nonumber\\
	{}& -\tfrac{1}{2}S_{330}\left(3 \tau_{\lambda } \tau_{\mu } \tau_{\nu }- 3\left\{ \tau a a \right\} _{\lambda\mu\nu} -\left\{\tau\eta\right\}_{\lambda\mu\nu}\right) 
	\nonumber\\
	{}& -\tfrac{1}{2}S_{333}\left(5 a_{\lambda } a_{\mu } a_{\nu }- \left\{ \tau \tau a \right\} _{\lambda\mu\nu} +\left\{ a \eta\right\}_{\lambda\mu\nu}\right) \,,
	\nonumber\\
	W_{\lambda\mu\nu} =
	{}&\tfrac{1}{2}W_{300}\bigl(\bigl[\tfrac{3}{2} \tau_{\lambda} \tau_{\{\mu} -\tfrac{1}{2}\eta_{\lambda\{\mu}\bigr]a_{\nu\}}-a_{\lambda} \bigl[3\tau_{\mu} \tau_{\nu}-\eta_{\mu\nu}\bigr]\bigr)
	\nonumber\\
	{}&+W_{330}\bigl(\bigl[\tfrac{3}{2} a_{\lambda} a_{\{\mu} +\tfrac{1}{2}\eta_{\lambda\{\mu}\bigr]\tau_{\nu\}}-\tau_{\lambda} \bigl[3a_{\mu} a_{\nu}+\eta_{\mu\nu}\bigr]\bigr)
	\nonumber\\
	{}&-\tfrac{1}{2}\epsilon^{3kl}W_{kl0}\,\tau^\alpha a^\beta\epsilon_{\alpha\beta\lambda\{\mu}\tau_{\nu\}}
	\nonumber\\
	{}&+\tfrac{1}{2}\epsilon^{3kl}W_{kl3}\,\tau^\alpha a^\beta\epsilon_{\alpha\beta\lambda\{\mu}a_{\nu\}}\,.
\end{align}
Expressions in curly brackets 
are understood to be symmetrized over even permutations of Lorentz indices,
e.g., 
$\{a\eta\}_{\lambda\mu\nu}=a_{\lambda}\eta_{\mu\nu}+a_{\mu}\eta_{\nu\lambda}+a_{\nu}\eta_{\lambda\mu}$,
while Lorentz indices enclosed between curly brackets 
are symmetrized in those indices, 
e.g., 
$A_{\{\mu}B_{\nu\}}\coloneqq A_{\mu}B_{\nu}+A_{\nu}B_{\mu}$.
As before, 
repeated $k,l \in \{1,2,3\}$ indices are summed over.

The momentum dependence of the torsion and nonmetricity coefficient vectors 
is organized by the nonrelativistic expansion defined in Eq.~\eqref{eq:FW_expansions}.
Explicit expressions for the expansion coefficients $\delta \bm{\xi_0}$, 
$\delta \bm{\xi_1}$, 
$\delta \bm{\xi_2}$, 
$\delta \bm{\zeta_0}$, 
$\delta \bm{\zeta_1}$, 
$\delta \bm{\zeta_2}$, 
and $\delta \bm{\zeta_3}$ 
for an arbitrary neutron-momentum direction $\bm{\hat{p}}=\bm{p}/|\bm{p}|$
can be derived from the Lagrangian~\rf{explL} for neutrons, 
$w=n$, 
using a generalized Foldy--Wouthuysen procedure~\cite{nonrel_limit}. 
For convenience we write $\hat{p}_z\coloneqq(\bm{\hat{p}})^3$ 
and $\bm{p}_{\varphi}\coloneqq\bm{\hat{z}}\times\bm{\hat{p}}$. 
The arrows below indicate the limit of axially propagating neutrons, 
$\bm{\hat{p}}\to\bm{\hat{z}}$, 
as relevant for the present analysis:
\begin{align}\label{xi_Def}
	\delta \bm{\xi_0}=
	{}&
	-\left[
	m \xi _5^{(5)} \epsilon ^{3kl} M^{0}{}_{kl} + \left(2 m \xi _9^{(5)}-\xi _4^{(4)}\right)A_3+\left(2 m \xi _8^{(5)}-\xi _2^{(4)}\right) T_3
	\right]\bm{\hat{z}}
	\nonumber\\
	\to
	{}&
	-\left[
	m \xi _5^{(5)} \epsilon ^{3kl} M^{0}{}_{kl} + \left(2 m \xi _9^{(5)}-\xi _4^{(4)}\right)A_3+\left(2 m \xi _8^{(5)}-\xi _2^{(4)}\right) T_3
	\right]\bm{\hat{z}}\,,
	\nonumber\\
	\delta \bm{\xi_1}=
	{}&
	\hat{p}_z\left[
	\tfrac{3}{2} m \xi_5^{(5)} \epsilon ^{3kl} M^{3}{}_{kl}
	\right]\bm{\hat{z}}
	-\left[
	\left(2 m \xi _9^{(5)}-\xi _4^{(4)}\right)A_0+\left(2 m \xi _8^{(5)}-\xi _2^{(4)}\right)T_0+\tfrac{1}{2} m \xi_5^{(5)} \epsilon ^{3kl} M^{3}{}_{kl}
	\right]\bm{\hat{p}}
	-3\left[
	m \xi _5^{(5)} M^{0}{}_{30}
	\right]\bm{p_\varphi}\,
	\nonumber\\
	\to 
	{}&
	\left[
	m \xi_5^{(5)} \epsilon ^{3kl} M^{3}{}_{kl}-\left(2 m \xi _9^{(5)}-\xi _4^{(4)}\right)A_0-\left(2 m \xi _8^{(5)}-\xi _2^{(4)}\right)T_0 
	\right]\bm{\hat{z}}\,,
	\nonumber\\
	\delta \bm{\xi_2}=
	{}&
	\tfrac{1}{2}\hat{p}_z\left[
	\left(\xi _4^{(4)}-2 m \xi _9^{(5)}\right) A_3+\left(
	\xi _2^{(4)}-2 m \xi _8^{(5)}\right) T_3+2m \xi _5^{(5)} \epsilon ^{3kl} M^{0}{}_ {kl}
	\right]\bm{\hat{p}}
	+3 m \hat{p}_z\left[\xi_5^{(5)} M^{3}{}_{30}\right]\bm{p_{\varphi}}
	\nonumber\\
	{}&-\tfrac{1}{2} \left[
	\left(\xi _4^{(4)}-2 m \xi _9^{(5)}\right) A_3+\left(
	\xi _2^{(4)}-2 m \xi _8^{(5)}\right) T_3+2m \xi _5^{(5)} \epsilon ^{3kl} M^{0}{}_ {kl}
	\right]\bm{\hat{z}}
	\nonumber\\
	\to 
	{}& 0\,,
\end{align}
and
\begin{align}\label{zeta_Def}
	\delta \bm{\zeta_0}=
	{}&
	\left[
	\tfrac{1}{2} \zeta _2^{(6)} m^2 S_{300}+\tfrac{1}{2} \zeta _5^{(5)} m \epsilon^{3jk}W_{jk0}
	+\left(\zeta _2^{(4)}-\zeta _9^{(5)} m\right) (N_1)_3+\left(\zeta _4^{(4)}-\zeta _{10}^{(5)} m\right) (N_2)_3 
	\right]\bm{\hat{z}},		
	\nonumber\\
	\to 
	{}&
	\left[
	\tfrac{1}{2} \zeta _2^{(6)} m^2 S_{300}+\tfrac{1}{2} \zeta _5^{(5)} m \epsilon^{3jk}W_{jk0}
	+\left(\zeta _2^{(4)}-\zeta _9^{(5)} m\right) (N_1)_3+\left(\zeta _4^{(4)}-\zeta _{10}^{(5)} m\right) (N_2)_3 
	\right]\bm{\hat{z}},	
	\nonumber\\
	\delta \bm{\zeta_1}=
	{}&
	\left[(N_1)_0 \left(\zeta _2^{(4)}-\zeta _9^{(5)} m\right)+(N_2)_0 \left(\zeta _4^{(4)}-\zeta _{10}^{(5)} m\right)+\tfrac{1}{2} \zeta _2^{(6)} m^2 \left(2 S_{000}-S_{330}\right)-\tfrac{3}{4} \zeta _6^{(5)}m W_{330}-\tfrac{1}{4}\zeta _5^{(5)} m\epsilon^{3jk}W_{jk3}\right]\bm{\hat{p}}
	\nonumber\\
	{}&+\tfrac{3}{8} m  \left[3 \zeta _5^{(5)} W_{300}-2 \zeta _6^{(5)} \epsilon^{3jk}W_{jk0}\right]\bm{p_{\varphi }}
	-\tfrac{1}{4} m \hat{p}_z \left[2 \zeta _2^{(6)} m \left(S_{000}-3 S_{330}\right)-3 \zeta_6^{(5)} W_{330}-3 \zeta _5^{(5)} \epsilon^{3jk}W_{jk3}\right] \bm{\hat{z}} \,,
	\nonumber\\
	\to
	{}&
	\left[(N_1)_0 \left(\zeta _2^{(4)}-\zeta _9^{(5)} m\right)+(N_2)_0 \left(\zeta _4^{(4)}-\zeta _{10}^{(5)} m\right)+\tfrac{1}{2} \zeta _2^{(6)} m^2 \left(S_{000}+2 S_{330}\right)+\tfrac{1}{2}m \zeta _5^{(5)} \epsilon^{3jk}W_{jk3}\right] \bm{\hat{z}} \,,		
	\nonumber\\
	\delta \bm{\zeta_2}=
	{}&
	\tfrac{1}{2} \hat{p}_z \left[\tfrac{7}{2} \zeta _2^{(6)} m^2 S_{300}-\zeta _2^{(6)} m^2 S_{333}- \zeta _5^{(5)} m
	\epsilon^{3jk}W_{jk0}+\left(\zeta _2^{(4)}-\zeta _9^{(5)} m\right) (N_1)_3+\left(\zeta _4^{(4)}-\zeta _{10}^{(5)} m\right) (N_2)_3\right]\bm{\hat{p}}
	\nonumber\\
	{}&+\tfrac{9}{4} m \hat{p}_z \left[\zeta _5^{(5)} W_{330}\right]\bm{p_\varphi}
	+\tfrac{1}{8} \left[2\zeta _2^{(6)} m^2 \left(-1+5\hat{p}_z{}^2\right) S_{333}+ 2\zeta _2^{(6)} m^2 \left(2-3 \hat{p}_z{}^2\right) S_{300}+4 \zeta _5^{(5)}m \epsilon^{3jk}W_{jk0}\right.
	\nonumber\\
	{}&\left.-4\left(\zeta _2^{(4)}-\zeta _9^{(5)} m\right) (N_1)_3-4\left(\zeta _4^{(4)}-\zeta _{10}^{(5)} m\right) (N_2)_3\right]\bm{\hat{z}}
	\nonumber\\
	\to 
	{}&
	\left[\tfrac{1}{2}\zeta _2^{(6)} m^2 \left(3 S_{300}+ S_{333}\right)\right] \bm{\hat{z}} \,,		
	\nonumber\\
	\delta \bm{\zeta_3}=
	{}&
	\tfrac{1}{8} \left[
	2 m^2 \zeta _2^{(6)} \left(3-2 \hat{p}_z{}^2\right) S_{000} -4 m^2 \zeta _2^{(6)} \left(1-3 \hat{p}_z{}^2\right) S_{330} +m \zeta _5^{(5)} \left(1-3
	\hat{p}_z{}^2\right) \epsilon ^{3kl}W_{kl3}+3 m \zeta _6^{(5)} \left(1-\hat{p}_z{}^2\right) W_{330}\right.
	\nonumber\\
	{}&\left. - 4 \left(\zeta _2^{(4)}-m \zeta _9^{(5)}\right) (N_1)_0-4 \left(\zeta_4^{(4)} - m \zeta _{10}^{(5)}\right) (N_2)_0 
	\right]\bm{\hat{p}}
	\nonumber\\
	\to 
	{}&
	\tfrac{1}{4}\left[	 m^2 \zeta _2^{(6)}  S_{000} +4 m^2 \zeta _2^{(6)}  S_{330} -m \zeta _5^{(5)} \epsilon ^{3kl}W_{kl3}
	- 2 \left(\zeta _2^{(4)}-m \zeta _9^{(5)}\right) (N_1)_0-2 \left(\zeta_4^{(4)} - m \zeta _{10}^{(5)}\right) (N_2)_0
	\right] \bm{\hat{z}} \,,
\end{align}
where 
we have dropped the $w=n$ label on the coupling constants $\xi$ and $\zeta$
for brevity.
In the above relations,
$\bm{\hat{z}}$ is the unit vector along the 3-axis, 
which is aligned with the electron polarization,
$\bm{p_\varphi}=\bm{\hat{z}}\times\bm{\hat{p}}$,
and $\hat{p}_z \coloneqq (\bm{\hat{p}})^3$ is the 3-component of $\bm{\hat{p}}$.
Note that a given torsion or nonmetricity component 
may contribute to $\delta {\bm \xi}$ or $\delta {\bm \zeta}$, 
respectively, 
at multiple orders in $\beta$. 
Note also 
that Eqs.~\rf{xi_Def} and ~\rf{zeta_Def} 
contain all 20 torsion and nonmetricity components
compatible with an axially symmetric source,
but that in the limit of axially propagating neutrons $\bm{\hat{p}}\to\bm{\hat{z}}$ 
sensitivity to the components 
$M^{0}{}_{30}$, $M^{3}{}_{30}$, $W_{300}$, and $W_{330}$ is lost.
Experimental access to these remaining four parameters 
requires neutron momenta 
not aligned with the net electron polarization.


\begin{thebibliography}{99}

	\bibitem{bl13}
	See, e.g.,
	M.~Blagojevi\'c and F.W.~Hehl, eds.,
	{\it Gauge Theories of Gravitation}, 
	Imperial College Press, London, 2013.
	
	\bibitem{Kibble1961}
	T.W.B.~Kibble,
	J.\ Math.\ Phys.\ {\bf 2}, 212 (1961).
	
	\bibitem{Sciama1964}
	D.W.~Sciama,
	Rev.\ Mod.\ Phys.\ {\bf 36}, 463 (1964).
	
	\bibitem{torsion}
	See, e.g.,
	F.W.~Hehl {\it et al.}, 
	Rev.\ Mod.\ Phys.\ {\bf 48}, 393 (1976);
	I.L.~Shapiro,
	Phys.\ Rep.\ {\bf 357}, 113 (2002);
	R.T.~Hammond,
	Rep.\ Prog.\ Phys.\ {\bf 65}, 599 (2002).
	
	\bibitem{ca22}
	E.~Cartan,
	C.R.\ Acad.\ Sci.\ (Paris) {\bf 174}, 593 (1922).
	
	\bibitem{Overview} 
	W.-T.~Ni,
	Rept.\ Prog.\ Phys.\ {\bf 73}, 056901 (2010).
	
	\bibitem{micro}
	N.F.~Ramsey, 
	Physica (Amsterdam) {\bf 96A}, 285 (1979);
	G.~Vasilakis, J.M.~Brown, T.W.~Kornack, and M.V.~Romalis,
	Phys.\ Rev.\ Lett.\ {\bf 103}, 261801 (2009);
	D.F.~Jackson Kimball, A.~Boyd, and D.~Budker,
	Phys.\ Rev.\ A {\bf 82}, 062714 (2010);
	E.G.~Adelberger and T.A.~Wagner,
	Phys.\ Rev.\ D {\bf 88}, 031101(R) (2013).
	
	\bibitem{SunSource} 
	W.-T.~Ni,
	Phys.\ Rev.\ D {\bf 19}, 2260 (1979).
	
	\bibitem{astro} 
	D.E.~Neville,
	Phys.\ Rev.\ D {\bf 21}, 2075 (1980);
	Phys.\ Rev.\ D {\bf 25}, 573 (1982);
	S.M.~Carroll and G.B.~Field,
	Phys.\ Rev.\ {\bf D 50}, 3867 (1994);
	G.~Raffelt,
	Phys.\ Rev.\ D {\bf 86}, 015001 (2012).
	
	\bibitem{Kaons} 
	S.~Mohanty and U.~Sarkar,
	Phys.\ Lett.\ B {\bf 433}, 424 (1998).
	
	\bibitem{CL97}
	C.~L\"ammerzahl,
	Phys.\ Lett.\ A {\bf 228}, 223 (1997).
	
	\bibitem{LHC}
	A.S.~Belyaev, I.L.~Shapiro, and M.A.B.~do Vale,
	Phys.\ Rev.\ D {\bf 75}, 034014 (2007).
	
	\bibitem{GravWave}
	S.~Capozziello, R.~Cianci, M.~De Laurentis, and S.~Vignolo,
	Eur.\ Phys.\ J.\ C {\bf 70}, 341 (2010).
	
	\bibitem{GProbeB} 
	Y.~Mao, M.~Tegmark, A.H.~Guth, and S.~Cabi,
	Phys.\ Rev.\ D {\bf 76}, 104029 (2007);
	see, however,
	E.E.~Flanagan and E.~Rosenthal,
	Phys.\ Rev.\ D {\bf 75}, 124016 (2007);
	F.W.~Hehl, Y.N.~Obukhov, and D.~Puetzfeld,
	Phys.\ Lett.\ A {\bf 377}, 1775 (2013).
	
	\bibitem{TorsionSME} 
	V.A.~Kosteleck\'y, N.~Russell, and J.D.~Tasson, 
	Phys.\ Rev.\ Lett. {\bf 100}, 111102 (2008).
	
	\bibitem{InMatter1}
	R.~Lehnert, W.M.~Snow, and H.~Yan,
	Phys.\ Lett.\ B {\bf 730}, 353 (2014);
	Erratum, Phys.\ Lett.\ B {\bf 744}, 415 (2015);
	H.~Yan \etal,
	Phys.\ Rev.\ Lett.\  {\bf 115}, 182001 (2015);
	A.N.~Ivanov and W.M.~Snow,
	Phys.\ Lett.\ B {\bf 764}, 186 (2017).
	
	\bibitem{InMatter2}
	B.~Arderucio Costa and Y.~Bonder,
	Phys.\ Lett.\ B {\bf 849}, 138431 (2024).
	
	\bibitem{PrefFrTests} 
	F.~Can\`e {\it et al.},
	Phys.\ Rev.\ Lett.\ {\bf 93}, 230801 (2004);
	B.R.~Heckel {\it et al.},
	Phys.\ Rev.\ Lett.\ {\bf 97}, 021603 (2006);
	B.R.~Heckel {\it et al.}, 
	Phys.\ Rev.\ D {\bf 78}, 092006 (2008).
	
	\bibitem{nm0}
	H.B.\ Weyl, 
	Sitzungsberichte der Preussischen Akademie der Wissenschaften {\bf 26}, 465 (1918).
	
	\bibitem{nm1} 
	J.~Foster, V.A.~Kosteleck\'y, and R.~Xu,
	Phys.\ Rev.\ D {\bf 95}, 084033 (2017).
	
	\bibitem{nm2}
	R.~Lehnert, W.M.~Snow, Z.~Xiao, and R.~Xu,
	Phys.\ Lett.\ B \textbf{772}, 865 (2017).
	
	\bibitem{Nico05b}
	J.S.~Nico {\it et al.},
	J.\ Res.\ Natl.\ Inst.\ Stand.\ Technol.\ {\bf 110}, 137 (2005).
	
	\bibitem{Dubbers11}
	D.~Dubbers and M.G.~Schmidt,
	Rev.\ Mod.\ Phys.\ {\bf 83}, 1111 (2011).
	
	\bibitem{Pignol:2015}
	G.~Pignol, 
	Int.\ J.\ Mod.\ Phys.\ A {\bf 30}, 1530048 (2015).
	
	\bibitem{Drummond:1979pp}
	I.T.~Drummond and S.J.~Hathrell,
	Phys.\ Rev.\ D \textbf{22}, 343 (1980).
	
	\bibitem{Bastianelli:2008cu}
	F.~Bastianelli, J.M.~Dávila, and C.~Schubert,
	JHEP \textbf{03}, 086 (2009).
	
	\bibitem{nonrel_limit}
	V.A.~Kosteleck\'y and C.D.~Lane,
	J.\ Math.\ Phys.\  {\bf 40}, 6245 (1999).
	
	\bibitem{ferri1}
	W.P.~Wolf,
	Rep.\ Prog.\ Phys.\ {\bf 24}, 212 (1961).
	
	\bibitem{ferri2}
	S.~Chikazumi,
	\textit{Physics of Ferromagnetism}, 2nd ed.\ (Oxford University Press, Oxford, 1997), p.~128.
	
	\bibitem{hughes_polarized_2025}
	C.D.~Hughes \textit{et al.},
	J.\ Magn.\ Magn.\ Mater.\ {\bf 629}, 173273 (2025).
	
	\bibitem{BaxterPrivate2026}
	D.V.~Baxter, private communication (2026).
	
	\bibitem{Mulkey2026}
	T.~Mulkey {\it et al.}, 
	Phys.\ Rev.\ Lett.\ {\bf 136}, 071801 (2026).
	
	\bibitem{iverson_flux_2024}
	E.B.~Iverson and M.L.~Crow, Jr.,
	\textit{Flux Measurements at CG-1D}, Technical Report ORNL/TM-2024/3701 (Oak Ridge National Laboratory, 2024).
	
	\bibitem{sme} 
	D.~Colladay and V.A.~Kosteleck\'y,
	Phys.\ Rev.\ D {\bf 55}, 6760 (1997);
	Phys.\ Rev.\ D {\bf 58}, 116002 (1998);
	V.A.~Kosteleck\'y,
	Phys.\ Rev.\ D {\bf 69}, 105009 (2004).
	
	\bibitem{DataTables}
	V.A.~Kosteleck\'y and N.~Russell, 
	{\it Data Tables for Lorentz and CPT Violation}, 
	2026 edition, arXiv:0801.0287v19.
	
\end{thebibliography}
\end{document}